\documentclass[prl,twocolumn,showpacs,amsmath,letter]{revtex4}
\usepackage{graphicx}
\newcommand{\Journal}[4]{#1 \textbf{#2}, #3 (#4)}

\begin{document}

\title{Current-Induced Dynamics in Almost Symmetric Magnetic Nanopillars}
\author{Weng Lee Lim}
\author{Andrew Higgins}
\author{Sergei Urazhdin}
\affiliation{Department of Physics, West Virginia University,
Morgantown, West Virginia 26506}


\begin{abstract}
Magnetic nanodevices usually include a free layer whose configuration can be changed by spin-polarized current via the spin transfer effect, and a fixed reference layer.  Here, we demonstrate that the roles of the free and the reference layers interchange over a small range of their relative thicknesses. Precession of both layers can be induced by spin transfer in symmetric devices, but the dynamics of one of the layers is rapidly suppressed in asymmetric devices. We interpret our results in terms of the dynamical coupling between magnetic layers due to spin transfer.
\end{abstract}
\maketitle

Spin torque (ST) exerted by spin-polarized current $I$ on a nanopatterned magnetic layer F$_1$ can change its  magnetic configuration or induce magnetic dynamics~\cite{cornellorig,cornellnature}. The polarization of $I$ is induced by another magnetic layer F$_2$, which is usually made much larger than F$_1$ to minimize the effects of ST on this layer. However, it is presently not known what makes a specific magnetic layer behave as a fixed polarizer or a free layer driven by ST. Significant effects of ST on both layers can be expected if their dimensions are similar, but the current-induced behaviors of such symmetric structures are not well understood. Magnetoresistance (MR) measurements of symmetric nanopillars indicated dynamics for both polarities of $I$, interpreted as independent excitations of each layer by the appropriate direction of $I$~\cite{tsoisymmetric}. However, calculations~\cite{stcoupling} and measurements of current-induced reversal statistics in symmetric nanopillars~\cite{wenglee} suggest that ST-induced coupling results in simultaneous incoherent dynamics of both layers.  

Significant effects of ST on the polarizing layer are also possible in asymmetric nanopillars~\cite{kentbipolar}.  Narrow spectroscopic peaks at large $I$ were attributed to the dynamics of the polarizer~\cite{largecurrent}, but their precise nature and the mechanism of excitation are not known.  Spectroscopy of current-induced dynamics in nanopillars with similar dimensions of F$_1$ and F$_2$ can potentially clarify these effects. Here, we report spectroscopic measurements of current-induced dynamics in almost symmetric nanopillars, where decoherence caused by ST-induced coupling between magnetic layers was reduced due to their different dynamical properties. We demonstrate coherent precession for both polarities of $I$ in symmetric devices. However, even slightly asymmetric devices exhibit a rapid suppression of precession for one of the current polarities, indicating significant effects of dynamical coupling between layers.

Multilayers Cu(60)F$_1$Cu(8)F$_2$Au(10)Cu(100), where thicknesses are in nm, were fabricated on oxidized silicon by a technique described elsewhere~\cite{limdynamics}. Only F$_2$=Py(5) (Py=Ni$_{80}$Fe$_{20}$) and part of Cu(8) spacer were patterned into an elliptical shape with approximate dimensions of $100$~nm~$\times$~$50$~nm, while F$_1$=Py($t_{Py}$) was left extended with dimensions of several $\mu$m. This sample geometry was verified by MR measurements, which showed negligible coupling between F$_1$ and F$_2$. All measurements were performed at room temperature (RT). The samples were contacted by coaxial microwave probes, which were connected through a bias tee to a current source, a lock-in amplifier, and a spectrum analyzer through a broadband amplifier. The microwave signals were adjusted for the frequency-dependent gain of the amplifier and losses in the cables and probes, determined with a microwave generator and a power meter. $I>0$ flowed from F$_1$ to F$_2$. The in-plane field ${\mathbf H}$ was rotated by $40^\circ$ with respect to the nanopillar easy axis, unless specified otherwise.  The rotation enabled spectroscopic detection of precession, and had an important effect on precession coherence~\cite{thadani}. 

The effects of ST on F$_1$ relative to F$_2$ were varied by changing its thickness $t_{Py}$.  We discuss current-induced behaviors of three samples with $t_{Py}=2,3$, and $4$~nm, labeled $Py2$, $Py3$, and $Py4$.
We shall see below that, because of the extended geometry of F$_1$, $Py3$ behaves as a symmetric sample with respect to the effects of current on F$_1$ and F$_2$, while the behaviors of $Py2$ and $Py4$ exhibit opposite asymmetries. The difference $\Delta R$ between resistance $R_{AP}$ in the antiparallel (AP) configuration and resistance $R_P$ in the parallel (P) configuration was $0.082$~$\Omega$, $0.124$~$\Omega$, and $0.134$~$\Omega$ for $Py2$, $Py3$, and $Py4$ respectively, consistent with stronger polarizing properties of increasingly thicker F$_1$.  The behaviors reported below were verified for at least two devices of each type.

\begin{figure}
\includegraphics[width=3.2in]{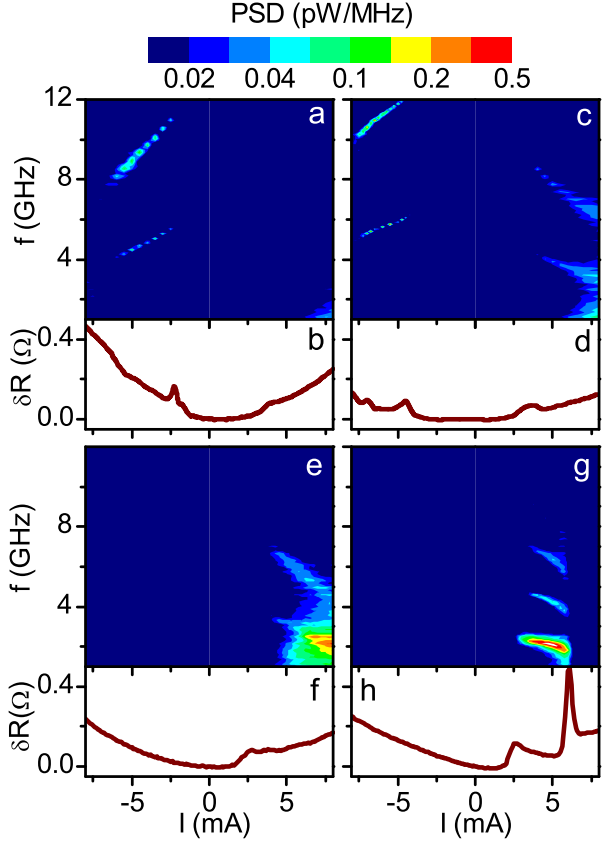}
\caption{\label{fig1} (Color online) (a) PSD {\it vs} frequency $f$ and $I$, and (b) $\delta R=dV/dI(I)-dV/dI(0)$ for sample $Py2$ at $H=600$~Oe. (c),(d) same as (a),(b), for $Py3$. (e),(f) same as (a),(b), for $Py4$, (g),(h) same as (e),(f), at $H=400$~Oe. }
\end{figure}

The power spectral density (PSD) data for sample $Py2$ exhibit two harmonically related spectral peaks  caused by current-induced precession at $I<0$ (Fig.~\ref{fig1}(a)). The onset of the microwave peaks above the measurement noise floor of $1$~fW/MHz up to $8$~GHz coincides with a small peak in differential resistance $dV/dI$ at $I^-_C=-2.3$~mA, Fig.~\ref{fig1}(b). A step in $dV/dI$ at $I^+_C=3.8$~mA corresponds to the onset of $1/f$ noise at $I>0$, characteristic of thermal transitions between different stable magnetic configurations of the bilayer.

In contrast to $Py2$, the spectroscopic data for sample $Py3$ exhibit precession peaks for both polarities of $I$ (Fig.~\ref{fig1}(c)). Their onsets coincide with bumps in $dV/dI$ at $I^-_C=-4.4$~mA
and $I^+_C=3.8$~mA (Fig.~\ref{fig1}(d)). The smallest full width at half maximum (FWHM) of the first harmonic at $I<0$ is $15$~MHz, similar to that seen in point contacts~\cite{nistmicrowaves}, but significantly smaller than in nanopatterned layers at RT~\cite{limdynamics}.  The first harmonic at $I>0$ has the smallest FWHM of $164$~MHz, and is superimposed on $1/f$ noise similar to that seen in $Py2$. The frequency of the first harmonic at the onset current $I^+_C$ is lower than the precession frequency at $I^-_C$. Based on these differences between the dynamical features at $I<0$ and $I>0$, we tentatively attribute the peaks at $I<0$ to precession of the extended F$_1$, and the peaks at $I>0$ to precession of the nanopatterned F$_2$. We generally expect that the current-induced dynamics of F$_2$ is localized in the area somewhat exceeding the dimensions of the point contact. The larger precession frequency of F$_1$ can be explained by the exchange interaction between the precessing part of F$_1$ and its static extended part.  The larger FWHM of the precession peaks at $I>0$ is likely caused  by thermally activated transitions between different configurations of F$_2$ stabilized by ST, also resulting in the $1/f$ noise~\cite{stcoupling}.

Sample $Py4$ shows precession peaks only at $I>0$, but no detectable dynamics at $I<0$ (Fig.~\ref{fig1}(e)). The $dV/dI$ curve (Fig.~\ref{fig1}(f)) has a step associated with the onset of current-induced dynamics only at $I>0$, but not at $I<0$, suggesting that dynamics is completely suppressed at $I<0$. It may be possible that a thicker F$_1$ has a precession onset outside of the measurement range of $-8$ to $8$~mA. However, assuming that $I^-_C$ is determined predominantly by the dimensions of F$_1$, one may expect that the ratio $I^-_C/I^+_C$ increases only by about a factor of $4/3$ for $Py4$ compared to $Py3$, giving $I^-_C \approx 4$~mA well within our measurement range.

Samples with F$_1$ much thicker than F$_2$ exhibited the smallest FWHM of precession peaks at $H$ just above $H_c$, quickly increasing as $H$ was increased~\cite{limdynamics}. Similarly, the $400$~Oe data for $Py4$ in Fig.~\ref{fig1}(g) exhibit smaller $1/f$ noise and narrower precession peaks than the $600$~Oe data, but the smallest FWHM of $165$~MHz is still about five times larger than in nanopillars with thick F$_1$. A peak in $dV/dI$ data (Fig.~\ref{fig1}(h)) at $I=6$~mA is correlated with abrupt termination of precession, which is caused by the transition of F$_2$ into a static AP orientation with respect to F$_1$~\cite{mystprl}.

\begin{figure}
\includegraphics[width=3.3in]{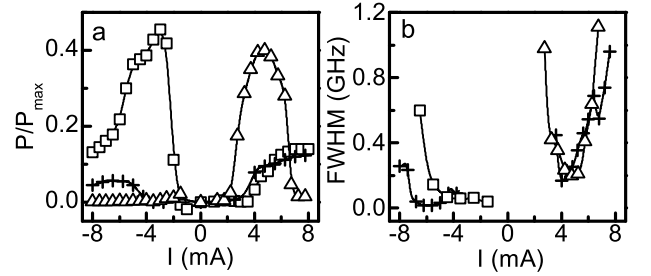}
\caption{\label{fig2} (a) Total microwave power emitted at $1-13.5$~GHz, normalized by the largest possible emitted power, calculated as described in the text. (b) FWHM of the first precession harmonic {\it vs} $I$.  Squares:  $Py2$ at $600$~Oe, crosses: $Py3$ at $600$~Oe, triangles: $Py4$ at $400$~Oe.}
\end{figure}

The magnetic dynamics can be further characterized by comparing the total microwave power emitted by the device to the largest possible power emitted by a hypothetical oscillation between AP and P configurations of the magnetic layers, $P_{max}=2I^2\Delta R^2R_L/(R+R_L)^2$. Here, $R$ is the average resistance during oscillation, and $R_L=50$~$\Omega$ is the input impedance of the amplifier. This formula was obtained by considering an equivalent electric circuit consisting of a dc current source, oscillating resistance representing the sample, and a load resistor $R_L$. Fig.~\ref{fig2}(a) shows the emitted power normalized by $P_{max}=10.8I^2$~pW, $27.7I^2$~pW, and $29.7I^2$~pW for $Py2$, $Py3$, and $Py4$, respectively, where $I$ is in mA. 

For $Py2$, the maximum relative power $P_r=P/P_{max}=0.45$ at $I=-3$~mA corresponds to the in-plane precession angle exceeding $90^\circ$, which is remarkable for an extended F$_1$. A significant reduction of $P_r$ at $I<-5.5$~mA is correlated with a dramatic increase of the FWHM of the precession peaks (Fig.~\ref{fig2}(b)). A similar correlation between the power and the FWHM of peaks is also observed in asymmetric samples with thick F$_1$, and is likely caused by the increasingly complex magnetic dynamics at large $I$~\cite{limdynamics}. In contrast, sample $Py3$ exhibits simultaneously a small $P_r<0.06$ at all $I<0$ (Fig.~\ref{fig2}(a)) and a FWHM of less than $30$~MHz at $-5$~mA$>I>-6.7$~mA  (Fig.~\ref{fig2}(b)). These behaviors of $Py3$ are transitional between a large power emitted by $Py2$ and a negligible power emitted by $Py4$ at $I<0$, suggesting that the dynamics of F$_1$ in $Py3$ is partly suppressed by the same mechanism that completely eliminates it in $Py4$.

The dependence of the emitted power on $t_{Py}$ at $I>0$ is opposite to that at $I<0$. Sample $Py4$ shows a maximum $P_r=0.41$ at $H=400$~Oe, while $Py2$ and $Py3$ show a significantly smaller $P_r<0.14$.  The maximum for $Py4$ at $600$~Oe is $P_r=0.28$ (not shown). The FWHM for $Py4$ first decreases with increasing $I<4.2$~mA. Its dramatic increase at larger $I$ is correlated with a decrease of $P_r$. The FWHM data for $Py3$ are surprisingly consistent with these behaviors, suggesting that the peak broadening is governed by the fundamental dynamical properties of the bilayers, rather than imperfections of individual devices.

\begin{figure}
\includegraphics[width=3.3in]{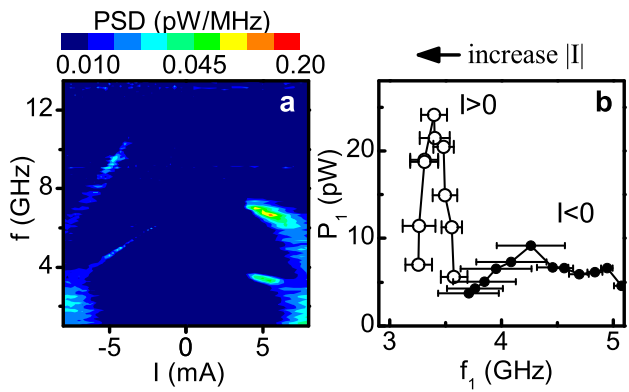}
\caption{\label{fig3} (Color online) (a) PSD for $Py2$ at $H=-600$~Oe, rotated by $55^\circ$ with respect to the nanopillar easy axis. (b) Integrated power $P_1$ in the first precession harmonic {\it vs} its peak frequency $f_1$ at $I<0$ (open symbols) and $I>0$ (solid symbols). Horizontal bars show FWHM. Arrow on top shows the direction of increasing $|I|$.}
\end{figure}

The dynamics of sample $Py2$ at $I>0$ became more coherent when the angle between ${\mathbf H}$ and the nanopillar easy axis was increased to $55^\circ$, as shown in Fig.~\ref{fig3}(a). This resulted in a decreased amplitude of peaks at $I<0$, as can be seen from the reduced color scale. We observe a similar reduction of the microwave power in all our samples exhibiting bipolar precession. The peaks in Fig.~\ref{fig3}(a) abruptly broaden at $I<-6$~mA, and the precession signal quickly decreases below the measurement noise floor. Figure~\ref{fig3}(b) shows that the frequency of precession at this point is correlated with the precession onset frequency of $3.6$~GHz at $I>0$. For $Py3$, broadening occurs at larger $I<-7$~mA (Fig.~\ref{fig2}(b)), which can be attributed to the higher precession frequency of F$_1$ in $Py3$ than in $Py2$ at the same $I$. We also see a similar broadening of precession at $I>5$~mA for $Py3$ and $Py4$ (Fig.~\ref{fig2}(b)), but its relation to the dynamical properties of F$_1$ is less clear.

To understand the implications of our results, we first  assume that the dynamics of F$_1$ at $I>0$ is negligible. ST acting on F$_2$ should result in its precession at $I>I^+_C>0$, which is determined only by the polarizing properties of F$_1$ and the magnetic moment of F$_2$. One can expect a modest increase of $I^+_C$ with decreasing $t_{Py}$ due to the weaker spin-polarizing properties of a thin F$_1$. Similarly, one may expect that the magnitude of the precession onset current $I^-_C$ for F$_1$ increases with increasing $t_{Py}$ due to the larger magnetic moment of F$_1$. Our results show that the precession of each layer rapidly disappears when its relative thickness is increased, instead of a gradual variation of the onset currents. This dramatic variation of the current-induced behaviors cannot be explained in terms of independent effects of ST on the two layers, and instead requires a self-consistent treatment of the {\it simultaneous} effects of ST on both layers.

Analysis of dynamics for magnetic layers with {\it similar} dynamical properties showed that ST-induced coupling can result in precession decoherence~\cite{stcoupling}, explaining the broadening of the precession peak at $I<0$ when its frequency became close to the onset frequency at $I>0$ (Fig.~\ref{fig3}). Coherent bipolar precession is not possible in structures with similar dynamical properties of the magnetic layers, because ST-induced coupling completely suppresses precession of the thicker layer.

To address the possibility of bipolar excitations in symmetric structures with {\it different} dynamical properties of layers, we develop a simple analytical model describing dynamics of two macrospins ${\mathbf S}_1$ and ${\mathbf S}_2$ coupled by ST. Here, the part of the extended layer F$_1$ in the point contact area is approximately treated as a macrospin ${\mathbf S}_1$ experiencing an additional effective field due to the exchange coupling to the extended film.  Linearizing the Landau-Lifshits-Gilbert-Slonczewski equation~\cite{slonczewski96} for small-amplitude precession of unit vectors ${\mathbf s}_{1,2}={\mathbf S}_{1,2}/|{\mathbf S}_{1,2}|$ around the equilibrium orientation along the $z$-axis, we obtain~\cite{stcoupling}
\begin{equation}
\label{xi} \frac{d\xi_{1,2}}{dt}=i\omega_{1,2}\xi_{1,2}-\Gamma_{1,2}\xi_{1,2}-Ik_{1,2}(\xi_1-\xi_2)
\end{equation}
where $\xi_{1,2}=s_{1x,2x}+is_{1y,2y}$ are complex coordinates representing the projections of the macrospins on the $xy$ plane. The first term on the right describes precession around the magnetic field, the second term describes relaxation, and the third term describes interaction between ${\mathbf s}_1$ and ${\mathbf s}_2$ due to ST. The coefficients $\omega_{1,2}$ depend on $\xi_{1,2}$ due to the demagnetizing fields of nanomagnets. By rescaling $s_x$, $s_y$ and $t$, this dependence can be transferred to $\Gamma_{1,2}$ and $k_{1,2}$. Assuming that the precession term is dominant, these coefficients can be replaced with their averages over the precession cycle.  Within these approximations, Eqs.~(\ref{xi}) with constant coefficients can be used to analyze the possibility of stable bipolar excitations in a symmetric bilayer.

Let's assume that $\xi_{1,2}$ precess at a common frequency $\omega$, with a precession phase offset $\phi$,  $\xi_1=A_1e^{i\omega t}$, and $\xi_2=A_2e^{i\omega t+i\phi}$. Equations~(\ref{xi}) then reduce to two complex algebraic relations
\begin{equation}
\label{omega} i\omega=i\omega_{1,2}-\Gamma_{1,2}\mp Ik_{1,2}[1-(re^{i\phi})^{\pm 1}]
\end{equation}
for the unknowns $\omega$, $\phi$, $r=A_2/A_1$, and $I$. The value of $I$ determined from these relations corresponds to the onset of the current-induced dynamics.  Assuming for simplicity $\Gamma_1=\Gamma_2$ and subtracting the two equations, we obtain $\cos\phi=(k_1+k_2)/(k_1r+k_2/r)$ from the real part, and $\sin\phi=(\omega_1-\omega_2)/(Ik_2/r-Ik_1r)$ from the imaginary part. For $I>0$, the dynamics of F$_2$ may be expected to dominate, so that $r\gg 1$. In this case, $\cos\phi\approx (k_1+k_2)/k_1r\ll 1$, and therefore $\sin\phi\approx -1$ for $\omega_1>\omega_2$, giving $r=(\omega_1-\omega_2)/Ik_1$. From Eqs. (\ref{omega}), $\omega\approx\omega_2$, and $I\approx\Gamma/k_2$. Stability analysis can be performed by assuming small deviations $\Delta r$ and $\Delta\phi$ of the respective variables from the stationary values determined above, and inserting them into Eqs. (\ref{xi}). The spectrum of the resulting linear differential system lies entirely in the left half of the complex plane, confirming the stability of this dynamical state.

From the expressions for $r$ and $I$ above, one can see that the condition $r\gg 1$ is satisfied for similar $k_1$ and $k_2$ only if $|\omega_1-\omega_2|\gg \Gamma$, i.e. when the dynamical properties of the magnetic layers are sufficiently different. For $|\omega_1-\omega_2|\le \Gamma$, the dynamics must instead involve precession of both layers with a similar amplitude~\cite{stcoupling}. Analysis similar to that given above also yields a stable dynamical state with $r\ll 1$ at $I<0$, as long as $\omega_1$ and $\omega_2$ are sufficiently different. In contrast, assuming $r\ll 1$ at $I>0$ or $r\gg 1$ at $I<0$, we obtain a dynamical state which is not stable with respect to small perturbations of $r$ and $\phi$.

Our analysis supports the possibility of bipolar excitations in symmetric bilayers with different dynamical properties of the layers. The model does not limit the relative thicknesses of the two layers, which  seems to be inconsistent with the rapid suppression of bipolar dynamics in asymmetric samples. We propose two possible mechanisms for this suppression. Firstly, small-angle precession frequency $f\approx 8$~GHz for F$_2$ at $H=600$~Oe can be calculated using Kittel formula. The onset frequency of about $4$~GHz in our data is rather associated with large-amplitude clamshell precession. Small-angle elliptical precession is not seen due to the small amplitude and the large width of the peaks at the precession onset (see Fig.~\ref{fig2}(b)). Because of this significant decrease of frequency with amplitude, F$_2$ can behave as a highly nonlinear oscillator whose frequency can autotune into resonance with F$_1$, suppressing the precession of the latter as long as the amplitude of precession of F$_2$ is not limited by its larger thickness. A similar argument can be used to explain the suppression of precession of F$_2$ by the dynamics of F$_1$. Secondly, for any precession frequency of F$_2$, there is an inhomogeneous dynamical mode of extended F$_1$ at the same frequency, which may be critical for suppression of the precession of F$_2$ due to the dynamical ST-induced coupling. It is not clear if a similar argument can be made for the suppression of the dynamics of F$_1$. A more detailed analysis explicitly including the nonlinearity of oscillations and inhomogeneous excitations may be needed to clarify these behaviors.

In conclusion, we have demonstrated bipolar current-induced precession in symmetric magnetic nanopillars with different dynamical properties of the two layers. Bipolar precession is rapidly suppressed in asymmetric devices.  This behavior is interpreted in terms of the dynamical coupling between magnetic layers induced by spin transfer. Our results suggest that coherent precession in nanomagnetic devices can be most efficiently achieved with significantly different dynamical properties of the magnetic layers. Only a small asymmetry between their dimensions is then needed to achieve strongly asymmetric behaviors of the bilayer.

We thank Lidiya Novozhilova for helpful discussions. This work was supported by the NSF Grant DMR-0747609 and a Cottrell Scholarship from the Research Corporation.

\end{document}